# Mediated Asymmetric Semi-Quantum Key Distribution

Yi-Fan Yang[1] and Tzonelih Hwang[*]

[1, *] *Department of Computer Science and Information Engineering, National Cheng Kung University, No. 1, University Rd., Tainan City, 70101, Taiwan, R.O.C.*

[1] ncku.yf.yang@gmail.com

[*] hwangtl@csie.ncku.edu.tw (corresponding author)

## Abstract

This study proposes a new mediated asymmetric semi-quantum key distribution (MASQKD) protocol. With the help of a dishonest third party, two classical participants, who have only limited **asymmetric** quantum capabilities, can share a secret key with each other. The proposed protocol is shown to be immune to several well-known attacks. Furthermore, an improved MASQKD protocol is proposed in which the quantum capabilities of one participant can be further reduced.

# 1 Introduction

The first quantum key distribution (QKD) protocol, which allows two participants to share a secret key with each other by using the properties of quantum mechanics, was proposed by Bennett and Brassard [1] in 1984. Subsequently, various QKD protocols have been proposed [2-6]. However, participants in these QKD protocols are assumed to have unlimited quantum capabilities, i.e., participants need to have various quantum devices, which are practically very expensive to implement the corresponding operations on.

To reduce participants' burden of making QKD protocols more practical, a semi-quantum key distribution (SQKD) protocol was proposed by Boyer et al. in 2009 [7]. In their protocol, a participant who has limited quantum capabilities is called a classical participant. A classical participant can only perform three of the following quantum operations: (1) measure qubits in Z basis $\{|0\rangle, |1\rangle\}$, (2) prepare qubits in Z basis, (3) reorder qubits via delay line, and (4) reflect or send qubits without disturbance. Moreover, in 2013, another quantum operation, i.e., perform a unitary operation on a qubit, was also considered as one of the possible operations for a semi-quantum environment in the semi-quantum information splitting protocol by Nie et al [8].

Besides SQKD, which allowed a quantum participant and a classical participant to share a secret key, some mediated SQKD protocols [9, 10] have been proposed to allow two classical participants to share a secret key securely between each other with the help of a third party (TP). In 2015, a mediated SQKD protocol proposed by Krawec [9] allowed two classical participants, who can only perform quantum operations (1), (2), and (4), to share a secret key securely. Liu et al. in 2018 also proposed a mediated SQKD protocol [10] without invoking quantum measurements. Classical users in Liu et al.'s protocol are limited to have the quantum capabilities (2), (3), and (4). Note that the integrity of TPs in both Krawec and Liu et al.'s protocols are assumed to be dishonest, which means that TP can perform any possible attack [9, 10].

However, the classical participants in this type of protocol are assumed to have the same (symmetric) quantum capabilities. In other words, two classical participants with asymmetric quantum capabilities have no way of sharing a secret key between each other by using the previously mentioned protocols. More specifically, a classical user with capabilities (1), (2), and (4) can only share a secret key with another user whose capabilities includes (1), (2), and (4). If one of the corresponding devices is not available or the two communicants are with different applications and cannot have symmetric quantum capabilities, then these two users can no longer share a key with each other. Therefore, if a mediated SQKD (i.e., MASQKD) can be designed for classical participants with asymmetric quantum capabilities, then the protocol can be more flexible and practical.

In this study, we aim to propose the first MASQKD protocol with the help of a dishonest TP. Security analyses show that the proposed protocol can detect and be immune to several well-known attacks. Furthermore, the proposed MASQKD is further modified to allow one of the two participants to have even less quantum capabilities.

The rest of the paper is organized as follows: Section 2 proposes the MASQKD protocol. In Section 3, the security analyses of the proposed protocol are provided, and the comparison between our protocol and other mediated SQKD protocols is discussed in Section 4. In Section 5, an improved MASQKD protocol is proposed. Finally, we conclude this study in Section 6.

## 2 The Proposed Protocol

This section describes the proposed MASQKD protocol. In the proposed protocol, two classical participants Alice and Bob, who have asymmetric quantum capabilities, want to share a secret key with the help of a dishonest TP. We assume that the quantum channels between the TP and Alice and between Alice and Bob in the proposed protocol are ideal (i.e., non-lossy and noiseless). An authenticated classical channel [1, 11-13] is connected between Alice and Bob.

In terms of classical participants' capabilities, one of the participants, i.e., Alice, owns capabilities including measuring and preparing qubits in Z basis $\{|0\rangle, |1\rangle\}$ and reflecting or sending qubits via quantum channels. On the other hand, the other participant, i.e., Bob, needs to perform Hadamard operation [14] on qubits and measure the qubits in Z basis, where the Hadamard operation is defined as follows:

$$H = \frac{1}{\sqrt{2}}\begin{pmatrix} 1 & 1 \\ 1 & -1 \end{pmatrix}.$$

In the proposed MASQKD protocol, the TP only needs to prepare and send $|+\rangle = \frac{1}{\sqrt{2}}(|0\rangle+|1\rangle)$ to Alice. The steps of the proposed protocol are described as follows (also shown in Figure 1):

**Step 1.** TP prepares $N$ (= $8n$) qubits ($S = \{s_1, s_2, ..., s_{8n}\}$) and sends them to Alice one by one, where $s_i = |+\rangle$, $\forall i = 1, 2, ..., 8n$.

**Step 2.** Once Alice receives qubit $s_i$ from the TP, two types of operations can be performed: (1) Alice can reflect $s_i$ to Bob directly, or (2) she can measure $s_i$ in Z basis, prepare a same quantum state, and send it to Bob (the operation is called measure-resend), where the reflected qubit or the qubit prepared by Alice is denoted as $s_i'$.

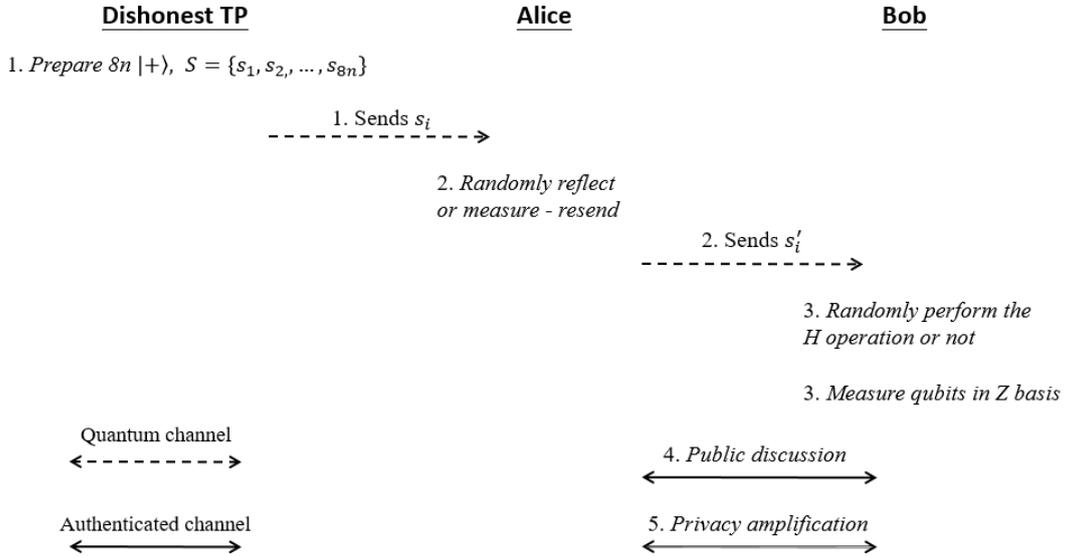

**Figure 1.** Proposed protocol.

**Step 3.** When Bob receives qubit $s_i'$, two types of operations are considered. He decides whether to perform the Hadamard operation before measuring $s_i'$ in the Z basis.

**Step 4.** In this step, Alice and Bob will start a discussion about the operations they had done in the previous step, i.e., for each qubit, Alice tells Bob her operation in Step 2, and Bob tells Alice whether he performs the Hadamard operation in Step 3. This checks the eavesdropper's and TP's integrity by discussing the measurement result via an authenticated channel. The possible cases are summarized as follows:

Case 1. If Alice reflects a qubit directly and Bob chooses to perform the Hadamard operation on it before measuring it in the Z basis, and if Bob's measurement result is not $|0\rangle$, then the qubit might have been compromised by an eavesdropper because the TP only generated $|+\rangle$ in the first step. Hence, if the error rate is higher than the predefined threshold, Alice and Bob will abort the current protocol.

Case 2. If Alice performs measure-resend on a qubit and Bob chooses not to perform the Hadamard operation before measuring the qubit in the Z basis, then Alice and Bob can share a common bit (the raw key), according to their operations. For example, if Alice's measurement result is $|0\rangle$ and Bob's measurement result is $|0\rangle$, a common bit will be shared between Alice and Bob.(refer to **Table 1.**)

Case 3. Except for the situations in Case 1 and 2, Alice and Bob will forgo the measurement results in all other situations (e.g., Alice performs measure-resend and Bob performs the Hadamard operation before measuring the qubit in the Z basis).

Table 1. Alice and Bob's sharing bit in Step4_Case 2.

| Alice's measurement result | Bob's measurement result | Sharing bit |
|---|---|---|
| $|0\rangle$ | $|0\rangle$ | 0 |

| $\|1\rangle$ | $\|1\rangle$ | 1 |

**Step 5.** After discussing all the results, Alice and Bob select half of the common bits and disclose them to check whether there is an eavesdropper and confirm the consistency of the common bits shared between them. The remaining bits are treated as the shared secret key after performing privacy amplification on them.

# 3 Security Analyses

In terms of a quantum key distribution protocol, collective attack is a very important class of attacks that most of the well-known attacks belong to it [15], such as modification attack and intercept-and-resend attack. Furthermore, collective attacks are considered to be the strongest joint attacks (the most general attack) [16]. Hence, the collective attacks analysis is given in this section. Moreover, besides the collective attacks analysis, to generate a truly secure key, participants need to obliterate the information which could have been obtained by Eve [16]. That is, Eve may just attack a few qubits to pass the detection processes. To solve this problem, the privacy amplification has been adapted in the above SQKD protocols where the raw key can be reduced to the truly secure key and the *key rate* is the most important parameter used for estimating the remaining length of the truly secure key in privacy amplification. Hence, the key rate analysis is shown in the section 3.2.

**3.1 Against Collective Attacks**

We first describe that the proposed protocol is immune to collective attacks, where the definition [15] is as follows:

(1) Each quantum system sent between users is attacked by Eve independently from others with the same strategy.

(2) Eve can keep her ancillary qubits until any later time, i.e., Eve can measure her ancillary qubits after obtaining some information coming from this attack.

To obtain useful information from the protocol by performing collective attacks, the attacker will entangle the initial quantum system with his prepared ancillary qubits and measure them later.

Since TP has more advantages than outsiders, any outsider attack can be resisted if the proposed protocol can resist attacks from a malicious TP. In this study, we want to show that TP cannot perform collective attacks on the proposed protocol to obtain useful information without disturbing the initial quantum system. Therefore, the malicious behavior will be detected by Alice and Bob.

There are two strategies of attacking the proposed protocol: (1) For the initial states sent from TP to Alice, TP performs collective attacks on them in Step 1. TP will measure his ancillary qubits to try to obtain useful information after Alice finishing her operations in Step 2; (2) TP performs collective attacks on the particles which are sent from Alice to Bob in Step 2. Similarly, TP tries to obtain information by measuring his ancillary qubits after Bob finishing his operations in Step 3. The discussions are as follows:

**Attacking Strategy 1:**

In this situation, TP makes the initial states entangle with ancillary qubits $E = \{|E_0\rangle, |E_1\rangle, ...\}$ by performing a $U$ operation on each qubit in Step 1, where $U$ satisfies $U*U = I$.

After Alice finishing her operations in Step 2, TP measures the ancillary qubits to obtain her measurement result and thus TP can infer the value of the shared bit by knowing the resultant state of the ancillary qubit.

First, we give the definition of the $U$ operation:

$$U|+\rangle \otimes |E_i\rangle = a_0 |+\rangle |e_0\rangle + a_1 |-\rangle |e_1\rangle$$
$$= \frac{1}{\sqrt{2}} \{|0\rangle (a_0 |e_0\rangle + a_1 |e_1\rangle) + |1\rangle (a_0 |e_0\rangle - a_1 |e_1\rangle)\} \quad (1)$$

, where the initial ancillary qubits are denoted as $|E_i\rangle$; $|a_0|^2 + |a_1|^2 = 1$; $|e_0\rangle$ and $|e_1\rangle$ are two different states which are distinguishable with each other by TP.

According to the equation (1), because TP can distinguish $|e_0\rangle$ and $|e_1\rangle$, he can also distinguish the states $a_0|e_0\rangle + a_1|e_1\rangle$ and $a_0|e_0\rangle - a_1|e_1\rangle$, hence TP can infer that whether Alice's measurement result is $|0\rangle$ or $|1\rangle$. However, this behavior will disturb the initial state. Alice and Bob can detect the error in Case 1 if TP sets $|a_1| \neq 0$.

That is, TP can only set $a_1|-\rangle|e_1\rangle$ as a zero vector to pass the discussion and obtain no useful information through this attacking strategy.

**Attacking Strategy 2:**

Another attacking strategy is to perform collective attacks on the qubit sent from Alice to Bob in Step 2. TP will measure the ancillary qubit to try to obtain Bob's measurement result after the original qubit is measured by Bob in Step 3. Also, the definitions of the U operation are given as follows:

$$U|0\rangle \otimes |E_i\rangle = a_0|0\rangle|e_0\rangle + a_1|1\rangle|e_1\rangle$$

$$U|1\rangle \otimes |E_i\rangle = b_0|0\rangle|f_0\rangle + b_1|1\rangle|f_1\rangle \qquad (2)$$

, where the initial ancillary qubit is denoted as $|E_i\rangle$; $|a_0|^2 + |a_1|^2 = 1$; $|b_0|^2 + |b_1|^2 = 1$; $|e_0\rangle$, $|e_1\rangle$, $|f_0\rangle$ and $|f_1\rangle$ are four different states which are distinguishable with each other by TP.

We can easily derive an equation below:

$$U|+\rangle \otimes |E_i\rangle = \frac{1}{\sqrt{2}}(U|0\rangle|E_i\rangle + U|1\rangle|E_i\rangle)$$

$$= \frac{1}{\sqrt{2}}(a_0|0\rangle|e_0\rangle + a_1|1\rangle|e_1\rangle + b_0|0\rangle|f_0\rangle + b_1|1\rangle|f_1\rangle) \quad (3)$$

$$= \frac{1}{2}\begin{pmatrix} |+\rangle \otimes (a_0|e_0\rangle + a_1|e_1\rangle + b_0|f_0\rangle + b_1|f_1\rangle) + \\ |-\rangle \otimes (a_0|e_0\rangle - a_1|e_1\rangle + b_0|f_0\rangle - b_1|f_1\rangle) \end{pmatrix}$$

This situation is similar to the Attacking Strategy 1, TP cannot pass the public discussion unless he set the term $a_0|e_0\rangle - a_1|e_1\rangle + b_0|f_0\rangle - b_1|f_1\rangle$ as a zero vector. Substituting it into the equation (3), we get the equation:

$$U|+\rangle \otimes |E_i\rangle = |+\rangle \otimes (a_0|e_0\rangle + b_0|f_0\rangle)$$

, which means $|a_0|^2 + |b_0|^2 = 1$. With this result, we can derive $|a_0| = |b_1|$; $|a_1| = |b_0|$, and indicate that $a_0|e_0\rangle = b_1|f_1\rangle$; $a_1|e_1\rangle = b_0|f_0\rangle$. Therefore, whether Alice sends $|0\rangle$ or $|1\rangle$ to Bob in Step 2, TP cannot distinguish $|0\rangle$ from $|1\rangle$ by measuring the ancillary qubit (refer to equation 2); that is, TP can obtain no useful information through this attacking strategy if he doesn't want to be detected by Alice and Bob.

### 3.2 Key Rate

In terms of the *key rate* (also called *secret fraction*), it is the actually meaningful quantity if the length of the key approaches infinity, and the definition is as follows:

$$r := \lim_{n \to \infty} \frac{l(n)}{n}$$

,where $n$ is the length of the raw key and $l(n)$ is the secure key size after error correction and privacy amplification. There is an elaborate analysis of key rate in Krawec's mediated SQKD protocol [9]. According to the equation of estimating the key rate [9]:

$$r \geq I(A:B) - \sup\{I(A:C)\}$$

, where A, B are two participants, C is the TP and $I(A:C)$ is the quantum mutual information between A and C. He first deals with the upper bound on $I(A:C)$ by considering the density matrix $\rho_{ABC}$ and its trace norm. Next, he computes the actually value of $I(A:B)$ and $\sup\{I(A:C)\}$. At the end of his analysis, the relation between key rate and error rate are discussed.

Because a detailed analysis of key rate is included in Krawec's protocol [9], to simplify our analysis, we want to approximately reduce the proposed protocol to his protocol to show that the two protocols have similar key rates. To begin with, let us provide a brief review of Krawec's protocol.

### 3.2.1 Krawec's Mediated SQKD Protocol

*A* and *B* are two classical participants who want to share a secret key with the help of a dishonest TP, *C*. The steps are as follows:

**Step 1.** *C* prepares Bell state $|\Phi^+\rangle = \frac{1}{\sqrt{2}}(|00\rangle + |11\rangle)$ and sends one particle to *A* and *B*, respectively.

**Step 2.** *A* (*B*) can (1) reflect, or (2) perform measure-resend on the received particle to *C*.

**Step 3.** *C* performs Bell measurement on the received particles and send message "$-1$" or "$+1$" to both *A* and *B*, which depends on the result is $|\Phi^-\rangle = \frac{1}{\sqrt{2}}(|00\rangle - |11\rangle)$ or not.

**Step 4.** *A* and *B* tell each other the operation they had done in **Step 2**. The possible cases are summarized as follows:

Case 1. If they both choose to reflect the qubit, then "$+1$" is the excepted message and "$-1$" is treated as an error. The protocol will be aborted if the error rate is higher than the predefined threshold.

Case 2. If they both perform measure-resend, then they can share a common bit if the message is "$-1$" and disregard their measurement if the message is "$+1$".

Case 3. Except for the situations in Case 1 and 2, Alice and Bob will forgo the measurement results in all other situations.

**Step 5.** *A* and *B* disclose some measurement results and check with each other. The remaining bits are treated as the shared secret key after performing privacy amplification on them.

### 3.2.2 Reduce to Krawec's Protocol

In order to finish this work, the proposed protocol's transmission path can be transformed into a modified one.

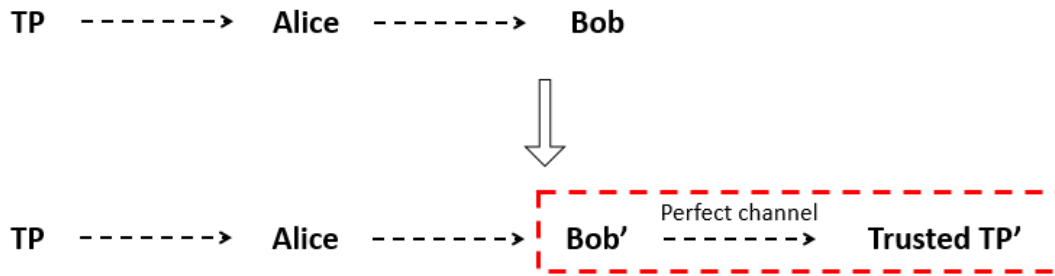

**Figure 2.** Transformation of the proposed protocol's transmission path

As shown in the **Figure 2**, Bob can be regarded as two characters, which are $Bob'$ and a trusted $TP'$. Note that a perfect channel without any disturbance is connected between Bob' and $TP'$ due to the fact that they are actually the same person. Therefore, Bob performs the Hadamard operation and then measures the qubit in Z basis also be regarded as $Bob'$ reflects the qubit directly to $TP'$ and TP' measures it in X basis.

Next, we are going to discuss all different cases of the two protocols.

Case 1. In the proposed protocol, the initial quantum resource is sent from TP to Alice, be reflected to $Bob'$, be reflected to $TP'$ and be measured in X basis by the $TP'$. Because $Bob'$ does not disturb the qubit (he reflect the qubit directly to TP'), the resource's transmission path is similar to the particle in Krawec's Protocol which is sent from *C* (the TP) to *A* and be reflected to *C*. For the same reason, since Alice reflects the qubit to $Bob'$ directly, the resource can be reduced to the particle in Krawec's Protocol which is sent from *C* (the TP) to *B* and be reflected to *C*. Note that the intentions in both protocols are to estimate the statistics of the quantum channels by checking the measurement result by the third party.

Case 2. In the proposed protocol, the initial quantum resource is sent from TP to Alice, be performed measure-resend by Alice and $Bob'$ respectively and afterwards, $TP'$ measures it in Z basis. Because $Bob'$ performs measure-resend on the qubit which is in Z basis, the qubits sent to $Bob'$ and $TP'$ should have the same quantum state. Therefore, the resource's transmission path is similar to the particle in Krawec's Protocol which is sent from $C$ (the TP) to $A$ and be performed measure-resend to $C$. Similarly, the resource can be reduced to the particle in Krawec's Protocol which is sent from $C$ (the TP) to $B$ and be performed measure-resend to $C$. After some iterations, the participants in both protocols can share some common bits (according to participants' measurement result). They will then disclose some selected bits and the secret key is generated after the privacy amplification.

Case 3. In both protocols, the classical users will forgo the measurement results except for the situations in Case 1 and 2.

Therefore, according to the above-mentioned analyses, the proposed protocol has the similar key rates with Krawec's protocol [9].

## 4 Comparison

In this section, we compare the proposed protocol with other mediated SQKD protocols, i.e., Krawec's [9] and Liu et al.'s protocols [10]. The comparison results are also shown in **Table 2**.

In both Krawec's and Liu et al.'s protocols, classical participants are limited to have the same quantum capabilities. As distinct from these protocols, two classical participants with asymmetric quantum capabilities can share a secret key with a dishonest TP in the proposed protocol.

In the proposed protocol, TP only needs to prepare single qubits in the X basis instead of preparing Bell states and does not require any quantum measurement device unlike in Krawec's and Liu et al.'s protocols. This obviously makes the protocol more practical. The qubit efficiency of the proposed protocol is $\frac{1}{12}$, which is lower than that of Liu et al.'s protocol but higher than that of Krawec's protocol with qubit efficiency η = n/m, where the number of shared key bits is denoted as n and the number of total prepared qubits is denoted as m.

**Table 2.** Comparison of MASQKD and mediated SQKD protocols

|  | **Proposed protocol** | **Krawec's [9]** | **Liu et al.'s [10]** |
|---|---|---|---|
| **TP's quantum capabilities** | 1. Prepare the single qubits in X basis | 1. Perform Bell measurement<br>2. Prepare Bell states | 1. Perform Bell measurement<br>2. Prepare Bell states |
| **Classical participant's quantum capabilities** | 1. Prepare<br>2. Measure<br>3. Reflect | 1. Prepare<br>2. Measure<br>3. Reflect | 1. Prepare<br>2. Reflect<br>3. Reorder |
|  | 1. Unitary operation<br>2. Measure |  |  |
| **Quantum resource** | Single qubits | Bell states | Bell states, Single qubits |
| **Qubit efficiency** | $\frac{1}{12}$ | $\frac{1}{24}$ | $\frac{1}{8}$ |

# 5 An Improved MASQKD Protocol

To make the proposed MASQKD even more practical, we propose an improved MASQKD protocol in this section. Because the cost of owning a quantum measurement device is much higher than that of performing a unitary operation, in the improved protocol, Bob does not need to own a quantum measurement device. Instead, Bob needs to perform a unitary operation $\sigma_Z$ and reflect a qubit to the TP, where $\sigma_Z$ is defined as follows:

$$\sigma_Z = \begin{pmatrix} 1 & 0 \\ 0 & -1 \end{pmatrix}$$

The steps of this protocol are described as follows (also shown in Figure 3):

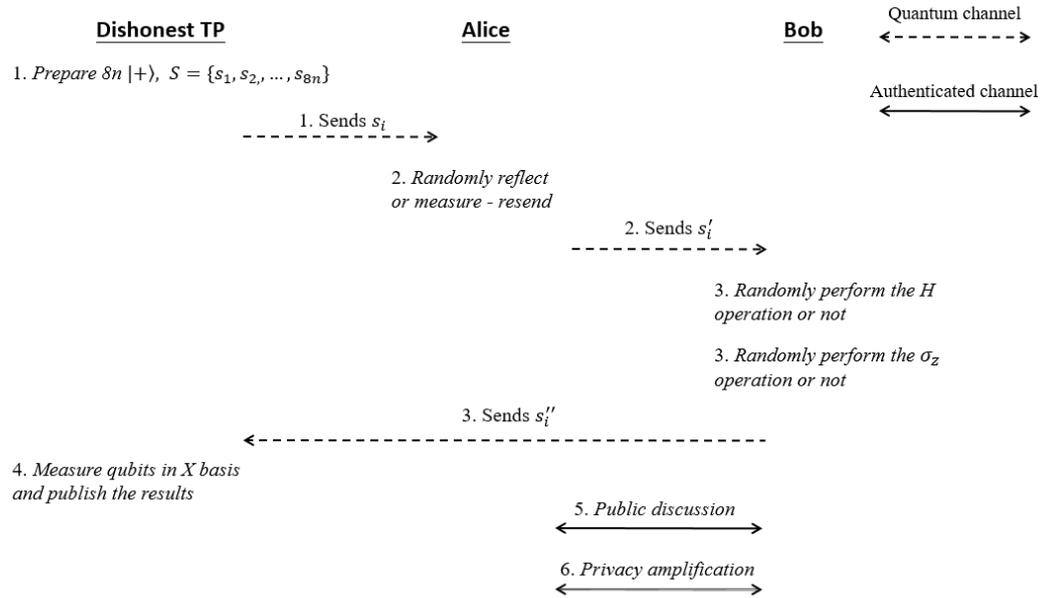

**Figure 3.** An improved protocol.

**Step 1** and **Step 2.** Follow the same process as Step 1 and 2 in Section 2.

**Step 3.** When Bob receives qubit $s_i'$ from Alice, Bob first decides whether to perform the Hadamard operation. Unlike the protocol in Section 2, Bob then chooses whether to perform the $\sigma_Z$ operation before reflecting the qubit to the TP. The reflected qubit is denoted as $s_i''$.

**Step 4.** The TP measures all the received qubits in the X basis and publishes all the measurement results.

**Step 5.** Similar to the previous protocol, Alice and Bob will start a discussion about the operations they performed in the previous step, i.e., for each qubit, Alice tells Bob her operation in Step 2 and Bob tells Alice whether he performs the Hadamard operation in Step 3. Note that Bob does not need to tell Alice whether he performs the $\sigma_Z$ operation. The possible cases are summarized as follows:

Case 1. If Alice reflects a qubit directly and Bob chooses not to perform the Hadamard operation on it before reflecting it to the TP, then TP's measurement result depends on Bob's decision in Step 3. That is, if Bob performs the $\sigma_Z$ operation, the result should be $\sigma_Z|+\rangle = |-\rangle$. Otherwise, the measurement result should be $|+\rangle$. If the corresponding result is not correct, the qubit might have been compromised by an eavesdropper. Hence, if the error rate is higher than the predefined threshold, Alice and Bob will abort the current protocol.

Case 2. If Alice performs measure-resend and Bob chooses to perform the Hadamard operation before reflecting the qubit to the TP, then Alice and Bob can share a common bit according to TP's measurement result and their operations. For example, suppose that TP's measurement result is $|+\rangle$. If Alice's measurement result is $|0\rangle$, she can infer that Bob did not perform the $\sigma_Z$ operation in Step 3. In addition, Bob can infer that Alice's measurement result is $|0\rangle$ in Step 2. A common bit is now shared between Alice and Bob.

Case 3. Except the situations in Case 1 and 2, Alice and Bob will forgo the measurement results in all other situations.

**Table 3.** Alice and Bob's sharing bit in Step5_Case 2.

| Alice's measurement result | Bob's operation | Expected measurement result | Shared classical bit |
|---|---|---|---|
| $\|0\rangle$ | H then $\sigma_Z$ | $\|-\rangle$ | 0 |
| $\|0\rangle$ | H | $\|+\rangle$ | 1 |
| $\|1\rangle$ | H then $\sigma_Z$ | $\|+\rangle$ | 0 |
| $\|1\rangle$ | H | $\|-\rangle$ | 1 |

**Step 6.** Follow the same process as Step 5 in Section 2. The remaining bits are treated as the shared key after Alice and Bob perform privacy amplification on the bits.

# 6 Conclusions

This study proposes a mediated asymmetric semi-quantum key distribution protocol. For the five quantum capabilities namely (1) measure qubits in Z basis $\{|0\rangle, |1\rangle\}$, (2) prepare qubits in Z basis, (3) reorder qubits via delay line, (4) reflect or send qubits without disturbance, and (5) perform a unitary operation on a qubit, the proposed protocol allows a classical user with quantum capabilities (1), (2), and (4) and the other classical user with quantum capabilities (1) and (5) to share a secret key with the help of a dishonest classical TP only with quantum capabilities (1) and (2). In terms of the security of the proposed protocol, an eavesdropper can be detected with an approximately 100 % detection probability if he/she tries to obtain useful information from the proposed protocol. Furthermore, an improved MASQKD protocol is proposed to reduce a user's quantum capabilities. In the improved MASQKD protocol, one of the two classical participants only performs unitary operations and reflects qubits without requiring the quantum measurement capability. It is indeed an interesting future research idea to design other MASQKD protocols with various combinations of simple quantum capabilities.

**Acknowledgements**

This research is supported by the Ministry of Science and Technology, Taiwan, R.O.C., under the Contract No. MOST 107-2221-E-006 -077 –; MOST 107-2627-E-006 -001 –; and MOST 107-2218-E-218 -004 –MY2.